\documentclass[sort&compress
    ,final            
  ,numberedheadings 
  ]
  {aipproc}

\layoutstyle{8x11single}
   \usepackage{showkeys}
    \usepackage{amssymb}
    \usepackage{bm}
    \usepackage{amsmath}

\newcommand\beq{\begin{equation}}
\newcommand\eeq{\end{equation}}
\newcommand\beqa{\begin{eqnarray}}
\newcommand\eeqa{\end{eqnarray}}

\begin{document}

\title{DSMC evaluation of the Navier--Stokes shear viscosity of a granular fluid
}

\author{Jos\'e Mar\'{\i}a Montanero}{
address={Departamento de Electr\'onica e Ingenier\'{\i}a Electromec\'anica, Universidad de Extremadura,  Badajoz, Spain}}
\author{Andr\'es Santos}{
  address={Departamento de F\'{\i}sica, Universidad de Extremadura, Badajoz, Spain}
}
\author{Vicente Garz\'o}{
  address={Departamento de F\'{\i}sica, Universidad de Extremadura, Badajoz, Spain}
}



\begin{abstract}
A method based on the simple shear flow modified by the introduction of a deterministic non-conservative force and a stochastic process is proposed to measure the Navier--Stokes shear viscosity in a granular fluid described by the Enskog equation. The method is implemented in DSMC simulations for a wide range of values of dissipation and density.
It is observed that, after a certain transient period, the system reaches a hydrodynamic stage which tends to the Navier--Stokes regime for long times. The results are compared with theoretical predictions obtained from the Chapman--Enskog method in the leading Sonine approximation, showing quite a good agreement, even for strong dissipation.

\end{abstract}

\maketitle


\section{Introduction}
\label{sec1}
Under the assumption of molecular chaos, the velocity distribution function of a \textit{dense} fluid of inelastic hard spheres obeys the Enskog kinetic equation, suitably modified to account for the collisional inelasticity through a constant coefficient of normal restitution $\alpha$ \cite{BDS97}. By an extension of the Chapman--Enskog method, the Navier--Stokes transport coefficients for a single fluid have been derived in the first Sonine approximation as nonlinear functions of the packing fraction $\phi$ and the coefficient of restitution \cite{GD99}. The shear viscosity for a dense binary mixture has also been evaluated within the same Sonine approximation \cite{GM03}.

Given that the first Sonine approximation is known to be rather accurate in the case of elastic hard spheres \cite{FK70}, a natural question is whether this approximation  is still reliable for finite inelasticity. This point is of interest since in the Sonine approximation the distribution function is expanded around the local \textit{equilibrium} distribution. However, a granular system is never in equilibrium \cite{G03}. In the undriven case, the uniform reference state corresponds to the so-called homogeneous cooling state (HCS), in which the time dependence only occurs through the temperature. The velocity distribution function of the HCS deviates from a Gaussian, as measured by a non-zero value of the fourth-cumulant and by an overpopulated high-energy tail \cite{vNE98}. This fact could cast some doubts about the accuracy of the Sonine approximation to compute the  transport  coefficients.

The shear viscosity  $\eta$ is perhaps the most widely studied transport coefficient in a granular fluid. For a low-density gas ($\phi\to 0$), Brey \textit{et al.} \cite{BRC96} measured $\eta$ from DSMC by analyzing the time decay of a weak transverse shear wave, and observed a qualitative good agreement with the Sonine prediction. However, while this method is physically quite elegant, it is not perhaps sufficiently efficient to measure $\eta$ accurately. More recently, the low-density Navier--Stokes transport coefficients of the HCS have been measured from DSMC by using the associated Green--Kubo formulas \cite{DB02}.
In the case of a system  \textit{heated} by the action of an external driving or thermostat, the associated shear viscosity (which is different from that of the HCS) has been computed from the Chapman--Enskog method in the first Sonine approximation and  from DSMC, for a dilute single gas \cite{GM02} as well as for a dense binary mixture \cite{GM03,MG02}.
On the other hand, to the best of our knowledge, the shear viscosity of the HCS for a \textit{dense} gas described by the Enskog equation has not been measured from DSMC before.
This issue is the main goal of this paper.
As a  second point,  we want to address the question of whether a granular fluid reaches a hydrodynamic regime that can be described by the Navier--Stokes equations, even far from the quasi-elastic limit. In fact, a granular fluid in an inhomogeneous \textit{steady} state is inherently non-Newtonian \cite{G03,SGD04}, so that a Newtonian regime requires \textit{unsteady} states sufficiently separated from the initial preparation.

In order to investigate the problem, we have carried out numerical simulations of the Enskog equation by means of an extension of the DSMC method \cite{MS96}  for $0.6\leq\alpha\leq 1$ and $0\leq \phi\leq 0.5$ and have considered the so-called simple shear flow state. To allow the granular fluid to approach a Newtonian regime, an external ``friction'' force with a negative friction coefficient is applied to compensate for the inelastic cooling, so that viscous heating produces a monotonic decrease of the Knudsen number (reduced shear rate) with time. In addition, to mimic the conditions appearing in the Chapman--Enskog method to Navier--Stokes order, a stochastic process is introduced, according to which every particle has a certain probability per unit time to have its velocity replaced by a new velocity sampled from the HCS distribution.

Our simulation results confirm that, regardless of its initial preparation, the fluid reaches after a few collisions per particle a hydrodynamic state whose temporal evolution is governed by that of the granular temperature. Moreover, when the shear viscosity is measured as the ratio between the shear stress and the shear rate for long times, it agrees reasonably well with the theoretical expressions derived in the first Sonine approximation \cite{GD99}, the deviations being comparable to or smaller than the ones in the elastic case.

\section{Modified simple shear flow}
The simple shear flow is macroscopically characterized by a constant density $n$, a uniform granular temperature $T$, and a linear velocity field $\mathbf{u}(\mathbf{r})=\mathsf{a}\cdot \mathbf{r}$, where the rate of strain tensor $\mathsf{a}$ is $a_{ij}=a\delta_{ix}\delta_{jy}$, $a$ being the constant shear rate \cite{GS03}.
At the level of the one-body distribution function, the simple shear flow becomes uniform in the Lagrangian frame of reference, i.e., $f(\mathbf{r},\mathbf{v},t)=f(\mathbf{V}(\mathbf{r}),t)$, where $\mathbf{V}(\mathbf{r})=\mathbf{v}-\mathbf{u}(\mathbf{r})$.
Consequently, the Enskog equation in this problem reads
\beq
\partial_t f-a V_y\frac{\partial}{\partial{V_x}} f
+{F}[f]= J[f,f].
\label{1}
\eeq
Here $J[f,f]$ is the Enskog operator, which in the simple shear flow is given by
\beq
J[f,f]= \sigma^2 \chi(n)\int d{\bf V}_1\int d\widehat{
{\bm{\sigma}}}\,
\Theta(\widehat{\bm{\sigma}}\cdot {\bf g})
(\widehat{\bm{\sigma}}\cdot {\bf g})
[\alpha^{-2}f({\bf V}')f({\bf V}_1'+\mathsf{a}\cdot\bm{\sigma})-f({\bf V})f({\bf V}_1-\mathsf{a}\cdot\bm{\sigma})],
\label{2}
\eeq
where $\sigma$ is the diameter of a sphere, $\chi(n)$ is the contact value of the pair distribution function, $\widehat{\bm{\sigma}}$ is a unit vector directed along the centers of the two colliding spheres at contact,  $\mathbf{g}=\mathbf{V}-\mathbf{V}_1$ is the relative velocity, $\alpha\leq 1$ is the coefficient of normal restitution,
$\bm{\sigma}\equiv\sigma \widehat{\bm{\sigma}}$, and the primes denote pre-collisional velocities, namely
${\bf V}'={\bf V}-\frac{1}{2}({1+\alpha^{-1}})
(\widehat{\bm{\sigma}}\cdot {\bf g})
\widehat{\bm{\sigma}}$ and ${\bf V}_1'={\bf V}_1+\frac{1}{2}({1+\alpha^{-1}})
(\widehat{\bm{\sigma}}\cdot {\bf g})
\widehat{\bm{\sigma}}$.
Finally, $F[f]$ in Eq.\ (\ref{1}) is an operator representing some type of external action on $f$, absent in the true simple shear flow problem, that will be chosen later on. This extra term is assumed to verify the conditions
\beq
\int d\mathbf{V} \{1,\mathbf{V}, V^2\}F[f]=\{0,\mathbf{0},-{3}n\gamma T/m\},
\label{4}
\eeq
so that it does not affect the mass and momentum conservation laws, but in general contributes to the energy balance equation through the ``heating'' rate $\gamma$.
The time evolution for the granular temperature is
\beq
\partial_t T=-\frac{2}{3n}a P_{xy}-\zeta T+\gamma T,
\label{5}
\eeq
where $P_{xy}=P_{xy}^k+P_{xy}^c$ is the \textit{total} shear stress and $\zeta$ is the cooling rate,
\beq
P_{ij}^k=\int d\mathbf{V} m V_i V_j f(\mathbf{V}),\quad P_{ij}^c=\frac{1+\alpha}{4}m\sigma^3\chi\int {d}\widehat{\bm{\sigma}}
\widehat{\sigma}_i\widehat{\sigma}_j\int {d}{\bf V}\int {d}{\bf V}_1
\Theta(\widehat{\bm{\sigma}}\cdot{\bf g})(\widehat{\bm{\sigma}}\cdot
{\bf g})^2 f({\bf V}+\mathsf{a}\cdot\bm{\sigma})f({\bf V}_1),
\label{3.6}
\end{equation}
\begin{equation}
\zeta=\frac{m\sigma^2}{12nT}(1-\alpha^2)\chi\int {d}\widehat{\bm{\sigma}}
\int {d}{\bf V}\int {d}{\bf V}_1
\Theta(\widehat{\bm{\sigma}}\cdot{\bf g})(\widehat{\bm{\sigma}}\cdot
{\bf g})^3 f({\bf V}+\mathsf{a}\cdot\bm{\sigma})f({\bf V}_1).
\label{3.7}
\end{equation}
The first term on the right-hand side of Eq.\ \eqref{5} represents viscous heating effects, the second term corresponds to the cooling due to the inelasticity of collisions, while the third term is the contribution due to the external action $F[f]$.

In the true simple shear flow, i.e., with  $F[f]=0$, the temperature evolution is governed by the competition between the viscous heating and the collisional cooling effects until a steady state  is reached when both effects cancel each other.
However, this steady state is inherently non-Newtonian \cite{SGD04}, so that the Navier--Stokes shear viscosity coefficient cannot be measured, even in the quasi-elastic limit.
The domain of validity of the Newtonian description is  restricted to small values of the Knudsen number $\text{Kn}=\lambda/\ell$, where $\lambda=1/\sqrt{2}\pi n\sigma^2\chi(n)$ is the mean free path and $\ell$ is the characteristic hydrodynamic length. In the present problem, the only hydrodynamic length is $\ell=\sqrt{2T/m}/a$. In the steady shear flow $\text{Kn}$ is a function of $\alpha$ that cannot be controlled independently. In order to reach asymptotically small values of $\text{Kn}$ for any value of $\alpha$ we need to avoid a steady state and have a monotonically increasing temperature. Consequently, we must modify the original shear flow problem by introducing an external driving mechanism, represented by $F[f]$, which exactly compensates for the collisional cooling term in Eq.\ \eqref{5}. Specifically, the heating rate introduced in Eq.\ \eqref{4} is chosen as $\gamma=\zeta$, so $\partial_t T=-(2/3n)aP_{xy}$.

So far, apart from the requirement  $\gamma=\zeta$, the explicit form of $F[f]$ remains still open. Here we will determine
$F[f]$ by requiring that the kinetic equation \eqref{1} to first order in $\text{Kn}$ be the same as the one found by applying the Chapman--Enskog method for states close to the (local) HCS \cite{GD99}. To that end we formally expand $f$ as
\beq
f(\mathbf{V})=f^{(0)}(\mathbf{V})+f^{(1)}(\mathbf{V})+\cdots,
\label{6}
\eeq
where $f^{(k)}(\mathbf{V})$ is of order $k$ in $\text{Kn}$. Moreover, the time dependence of $f$ occurs entirely through the temperature. Note that, by definition, $a\sim \text{Kn}$. On physical grounds, $P_{xy}$ is at least of first order in $\text{Kn}$. Therefore, given that $\gamma=\zeta$, $\partial_t T\sim \text{Kn}^2$ and so $\partial_t f$ can be neglected to first order. Inserting Eq.\ \eqref{6} into Eq.\ \eqref{1}, we get, through first order in $\text{Kn}$,
\beq
F[f^{(0)}]=J^{(0)}[f^{(0)},f^{(0)}],
\label{7}
\eeq
\beq
F[f^{(1)}]-a V_y\frac{\partial}{\partial V_x}f^{(0)} = -L[f^{(1)}]+a \Lambda_y\left[{\partial  f^{(0)}}/{\partial V_x}\right],
\label{8}
\eeq
where
\beq
J^{(0)}[X,Y]=
\sigma^2 \chi(n)\int d{\bf V}_1\int d\widehat{
{\bm{\sigma}}}\,
\Theta(\widehat{\bm{\sigma}}\cdot {\bf g})
(\widehat{\bm{\sigma}}\cdot {\bf g})
[\alpha^{-2}X({\bf V}')Y({\bf V}_1')-X({\bf V})Y({\bf V}_1)],
\label{9}
\eeq
\beq
L[X]=-J^{(0)}[f^{(0)},X]-J^{(0)}[X,f^{(0)}],
\label{10}
\eeq
\beq
\Lambda_i[X]=
\sigma^3 \chi(n)\int d{\bf V}_1\int d\widehat{
{\bm{\sigma}}}\,
\Theta(\widehat{\bm{\sigma}}\cdot {\bf g})
(\widehat{\bm{\sigma}}\cdot {\bf g})\widehat{\sigma}_i
[\alpha^{-2}f^{(0)}({\bf V}')X({\bf V}_1')+f^{(0)}({\bf V})X({\bf V}_1)].
\label{11}
\eeq

We require that $f^{(0)}$ be the HCS. This implies that \cite{vNE98}
\beq
F[f^{(0)}]=\frac{1}{2}\zeta^{(0)}\frac{\partial}{\partial\mathbf{V}}\cdot \left[\mathbf{V}f^{(0)}(\mathbf{V})\right],
\label{12}
\eeq
where $\zeta^{(0)}$ is the cooling rate in the HCS, which is obtained from Eq.\ \eqref{3.7} by setting $a=0$ and $f\rightarrow f^{(0)}$.
Next, we consider the linear integral equation \eqref{8} for $f^{(1)}$. This equation coincides with the one derived in Ref.\ \cite{GD99} from the general Chapman--Enskog method specialized to $\nabla n=\nabla T=\nabla\cdot \mathbf{u}=0$, provided that
\beq
F[f^{(1)}]=-\zeta^{(0)}T{\partial_T}f^{(1)}= \frac{1}{2}\zeta^{(0)}\frac{\partial}{\partial\mathbf{V}}\cdot \left[\mathbf{V}f^{(1)}(\mathbf{V})\right]+\frac{1}{2}\zeta^{(0)}f^{(1)}(\mathbf{V}),
\label{13}
\eeq
where in the last step we have taken into account that $\text{Kn}\propto T^{-1/2}$ and, by dimensional analysis, $f^{(1)}(\mathbf{V})=n  \text{v}_0^{-3}\Phi(\mathbf{V}/\text{v}_0)\text{Kn}$, where $\text{v}_0=\sqrt{2T/m}$ is the thermal velocity, $\Phi$ being a dimensionless function of the reduced velocity.
The simplest choice for $F[f]$ compatible with Eq.\ \eqref{4} (with $\gamma=\zeta$) and with Eqs.\ \eqref{12} and \eqref{13} is
\beq
F[f]=\frac{1}{2}\zeta\frac{\partial}{\partial\mathbf{V}}\cdot \left(\mathbf{V}f\right)+ \frac{1}{2}\zeta \left(f-f^{(0)}\right).
\label{14}
\eeq

In summary, our \textit{modified} simple shear flow problem is described by Eq.\ (\ref{1}) along with the choice \eqref{14}.
The first term on the right-hand side of Eq.\ \eqref{14} represents the effect of a deterministic non-conservative force of the form $\frac{1}{2}\zeta m\mathbf{V}$, which does work to compensate for the collisional energy loss. The shear viscosity of a granular fluid mixture when only this term is present in $F[f]$ has been determined from the Chapman--Enskog method and measured in DSMC simulations \cite{MG02,GM03}. However, this term does not suffice to get the Navier--Stokes shear viscosity of the HCS, but it must be supplemented by the second term on  the right-hand side of Eq.\ \eqref{14}. The latter term represents the action of a \textit{stochastic} process, whereby each particle has a probability per unit time equal to $\frac{1}{2}\zeta$ of changing its velocity by a new velocity sampled from the (local) HCS distribution $f^{(0)}$. When this stochastic term is moved to the right-hand side of the Enskog equation \eqref{1}, it can be interpreted  as a BGK-like relaxation term.

\section{DSMC results}
\begin{figure}[tbp]
\includegraphics[width=1.\columnwidth]{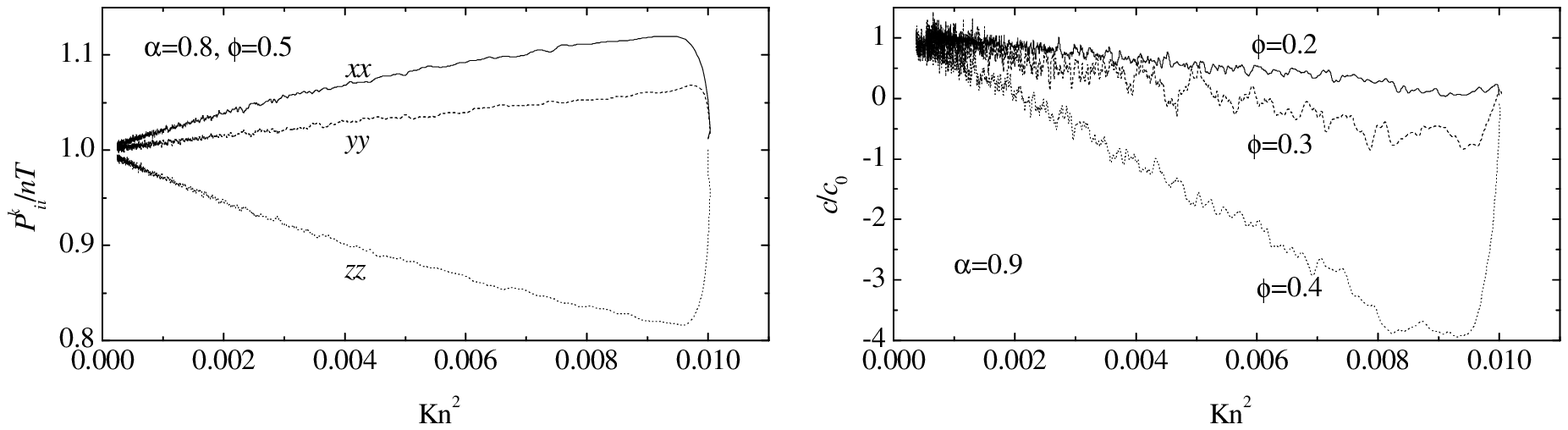}
\caption{Left panel: Parametric plot of the reduced normal stresses $P_{ii}^k/nT$ versus $\text{Kn}^2$ for $\alpha=0.8$ and $\phi=0.5$. Right panel: Parametric plot of the fourth cumulant $c$, relative to that of the HCS, versus $\text{Kn}^2$ for $\alpha=0.9$ and $\phi=0.2$, 0.3, and 0.4.
\label{fig1}}
\end{figure}
We have solved the Enskog equation \eqref{1} with the presence of $F[f]$ as given by Eq.\ \eqref{14} by means of an extension to the Enskog equation \cite{MS96} of Bird's DSMC method. Since the system is uniform in the Lagrangian frame, there is no need to split it into cells. In the implementation of $F[f]$ the cooling rate is self-consistently measured in the simulations. The stochastic term is simulated by randomly choosing during a small time step $\delta t$ a fraction $\frac{1}{2}\zeta \delta t$ of particles; the original velocity $\mathbf{V}_{\text{old}}$ of each one of those particles is replaced by a new velocity $\mathbf{V}_{\text{new}}=\sqrt{T}\mathbf{C}$, where $\mathbf{C}$ is the velocity of a particle in a reservoir kept at the HCS normalized to unit temperature.

The shear rate and the initial temperature are such that $\text{Kn}=0.1$ at $t=0$ for all density and coefficient of restitution. Besides, the initial velocity distribution is a Maxwellian. In the course of the simulations, the kinetic and collisional transfer contributions \eqref{3.6} to the pressure tensor are evaluated. {}From the shear stress $P_{xy}(t)$ the shear viscosity is measured as a function of time as $\eta(t)=-P_{xy}(t)/a$. Since the Knudsen number $\text{Kn}(t)\propto 1/\sqrt{T(t)}$ monotonically decreases with increasing time, the Navier--Stokes shear viscosity is identified  as $\eta(t)$ for long times.

As said in Section \ref{sec1}, our main objective is to compare the kinetic ($\eta^k$) and total ($\eta$) shear viscosity measured in the simulations with the expressions derived from the Chapman--Enskog method by using the first Sonine approximation $f^{(1)}\to -a(m\eta^k/nT^2)f_M V_xV_y$, $f_M$ being the Maxwellian distribution. The expressions are \cite{GD99}
\beq
\eta^k(\alpha,\phi)=\eta_0\frac{1-\frac{2}{5}\phi\chi(\phi)(1+\alpha)(1-3\alpha)}{\frac{1}{384}\chi(\phi)(1+\alpha)\left[16(13-\alpha)-3(4-3\alpha)c_0(\alpha)\right]},
\label{15}
\eeq
\beq
\eta(\alpha,\phi)=\eta^k(\alpha,\phi)\left[1+\frac{4}{5}\phi\chi(\phi)(1+\alpha)\right]+\eta_0
\frac{384}{25\pi}\phi^2\chi(\phi)(1+\alpha)\left[1-\frac{1}{32}c_0(\alpha)\right],
\label{16}
\eeq
where $\eta_0=(5/16\sigma^2)\sqrt{mT/\pi}$ is the shear viscosity of a dilute gas in the elastic limit, $\phi=\frac{\pi}{6}n\sigma^3$ is the packing fraction, and $c_0$ is the fourth cumulant of the HCS. A good estimate of $c_0$ is \cite{vNE98}
\beq
c_0(\alpha)=\frac{32(1-\alpha)(1-2\alpha^2)}{81-17\alpha+30\alpha^2(1-\alpha)}.
\label{17}
\eeq
In addition, we take the Carnahan--Starling approximation $\chi(\phi)=(1-\phi/2)/(1-\phi)^3$.

Figure \ref{fig1} shows the kinetic part of the normal stresses $P_{ii}^k(t)$, relative to $nT(t)$, and the fourth cumulant $c(t)$, relative to its HCS value $c_0$, as functions of the time-dependent Knudsen number $\text{Kn}(t)=\lambda/\ell(T)=a\sigma/12\phi\chi(\phi)\sqrt{T(t)/m}$, for some representative cases. Note that, as time grows, the Knudsen number $\text{Kn}$ monotonically decreases from its initial value $\text{Kn}=0.1$, behaving as $\text{Kn}(t)\sim t^{-1}$ for asymptotically long times. We observe that in the long-time limit, i.e., for $\text{Kn}\to 0$, the system tends to an isotropic state with a fourth cumulant $c$ equal to that of the HCS. This supports the expectation that the asymptotic state of our modified simple shear flow is the HCS, despite the fact that the temperature is increasing rather than decreasing, in agreement with Eq.\ \eqref{7}. It is worth remarking that, according to Fig.\ \ref{fig1}, after a short transient period the fluid reaches a \textit{hydrodynamic} regime where the normal stresses and the cumulant are linear functions of $\text{Kn}^2$ (Burnett-order effects).

\begin{figure}[tbp]
\includegraphics[width=.50\columnwidth]{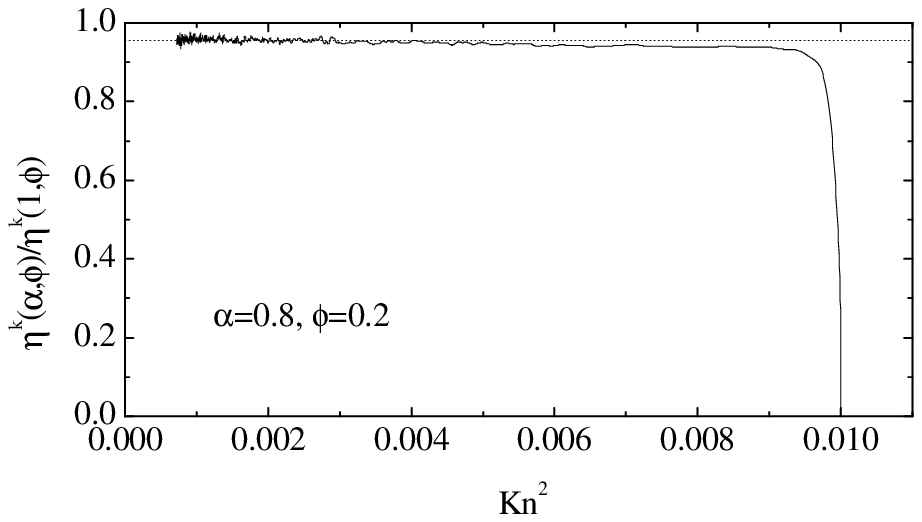}
\caption{Parametric plot of the reduced kinetic shear viscosity $\eta^k(\alpha,\phi)/\eta^k(1,\phi)$  versus $\text{Kn}^2$ for $\alpha=0.8$ and $\phi=0.2$. The dotted line represents the estimated Navier--Stokes value.
\label{fig1bis}}
\end{figure}

As an illustration of how the Navier--Stokes shear viscosity is evaluated from DSMC, Fig.\ \ref{fig1bis} shows the time-dependent kinetic shear viscosity $\eta^k(t)=-P_{xy}^k(t)/a$, relative to its (time-dependent) theoretical Sonine value in the elastic limit,
 as a function of $\text{Kn}^2(t)$ for the case $\alpha=0.8$, $\phi=0.2$. Figure  \ref{fig1bis} clearly shows that the ratio $\eta^k(\alpha,\phi)/\eta^k(1,\phi)$ reaches a plateau for long times (small Knudsen numbers) that can be identified as the Navier--Stokes value. The same procedure has been followed to measure the kinetic and collisional parts of the Navier--Stokes shear viscosity for different values of dissipation and density.
\begin{figure}[h!]
\includegraphics[width=1.\columnwidth]{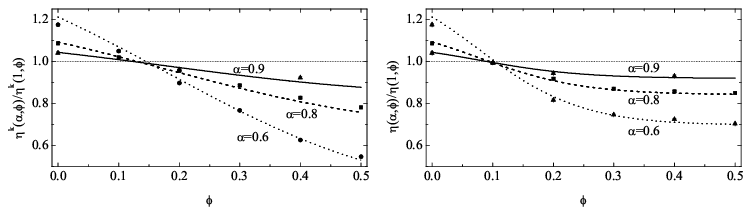}
\caption{Plots of the kinetic part of the shear viscosity (left panel) and of the total shear viscosity (right panel), relative to their respective theoretical values in the elastic limit in the first Sonine approximation, as  functions of the packing fraction for $\alpha=0.9$, 0.8, and 0.6. The symbols are simulation results, while the lines are the theoretical predictions in the first Sonine approximation.
\label{fig2}}
\end{figure}
\begin{figure}[h!]
\includegraphics[width=1.\columnwidth]{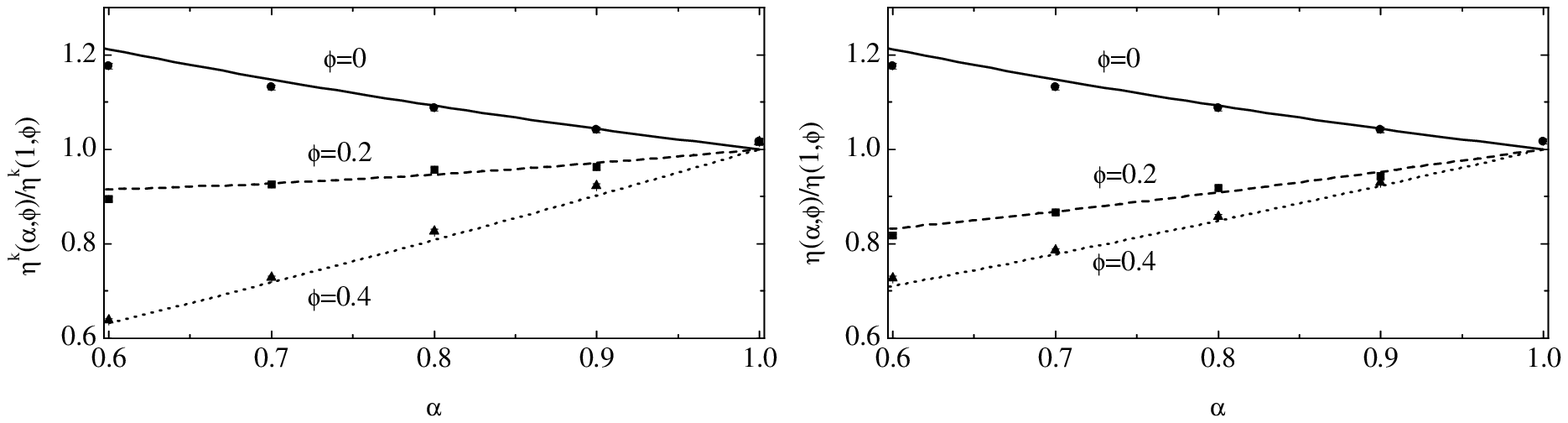}
\caption{Plots of the kinetic part of the shear viscosity (left panel) and of the total shear viscosity (right panel), relative to their respective theoretical values in the elastic limit in the first Sonine approximation, as  functions of the coefficient of restitution for $\phi=0$, 0.2, and 0.4. The symbols are simulation results, while the lines are the theoretical predictions in the first Sonine approximation.
\label{fig3}}
\end{figure}

The density dependence of $\eta^k$ and $\eta$ for three values of the coefficient of restitution is displayed in Fig.\ \ref{fig2}, while Fig.\ \ref{fig3} shows the influence of dissipation for three values of the packing fraction.
We observe a general good agreement between DSMC results and the predictions from the first Sonine approximation, even for strong dissipation and  high densities. Both theory and simulation show that, at a given value of $\alpha$, $\eta^k(\alpha,\phi)>\eta^k(1,\phi)$ if the packing fraction is smaller than a certain value $\phi_0^k(\alpha)$, while $\eta^k(\alpha,\phi)<\eta^k(1,\phi)$ if $\phi>\phi_0^k(\alpha)$. A similar behavior occurs for the total shear viscosity with a different value $\phi_0(\alpha)$. The influence of $\alpha$ on both $\phi_0^k$ and $\phi_0$ is rather weak; according to Eqs.\ \eqref{15}--\eqref{17}, $(\phi_0^k,\phi_0)= (0.12,0.09 )$, $(0.13,0.09 )$, and $(0.15,0.10 )$ for $\alpha=0.9$, 0.8, and 0.6, respectively.
As a consequence, while in a dilute granular gas ($\phi\lesssim 0.1$) the shear viscosity increases with inelasticity, the opposite happens for sufficiently dense fluids ($\phi\gtrsim 0.1$).

\section{Concluding remarks}
In this paper we have proposed a method to measure the Navier--Stokes shear viscosity of a moderately dense granular fluid described by the Enskog equation. The idea is to consider the simple shear flow (which is uniform in the Lagrangian frame), modified by the presence of a deterministic non-conservative force (which compensates for the collisional cooling) along with a stochastic BGK-like term. Under these conditions the Knudsen number of the problem decreases with increasing time, so that the system reaches a hydrodynamic Navier--Stokes regime for long times. This procedure allows one to evaluate the Navier--Stokes shear viscosity in an efficient way by means of the DSMC method. The simulation results have been compared with  predictions  from the Chapman--Enskog expansion in the first Sonine approximation \cite{GD99}.
The results show that the Sonine predictions compare quite well with the simulation data for the wide range of dissipation ($\alpha\geq 0.6$) and density ($\phi\leq 0.5$) explored. This agreement is significant, given that, in contrast to the elastic case, the reference state is not a (local) Maxwellian but the (local) HCS, which exhibits non-Maxwellian features, such as a non-zero fourth cumulant and an overpopulated high-velocity tail \cite{vNE98}. It is interesting to remark that the accuracy of the Sonine approximation found here is comparable to the one observed for elastic fluids, so that one could expect to improve the agreement by considering more terms in the Sonine polynomial expansion.
Finally, we want to stress that the system, after a short transient period, achieves a hydrodynamic stage prior to the Navier--Stokes regime. This stage could be used to study nonlinear transport properties (e.g., shear thinning and viscometric effects), although this issue is beyond the scope of this paper.


\begin{theacknowledgments}
Partial support from the Ministerio de
Educaci\'on y Ciencia
 (Spain) through Grants Nos.\ ESP2003-02859 (J.M.M.) and FIS2004-01399 (A.S. and V.G.) is gratefully acknowledged.
\end{theacknowledgments}


\bibliographystyle{aipproc}   

\begin{thebibliography}{99}


\bibitem{BDS97}
 Brey, J. J.,  Dufty, J. W., and  Santos, A., \textit{J. Stat. Phys.} \textbf{87}, 1051--1066 (1997).

\bibitem{GD99}
 Garz\'o, V., and  Dufty, J. W., \textit{Phys. Rev. E} \textbf{59}, 5895--5911 (1999).

\bibitem{GM03}
 Garz\'o, V., and  Montanero, J. M., \textit{Phys. Rev. E} \textbf{68}, 041302-1--17 (2003).

\bibitem{FK70}
 Ferziger, J., and  Kaper, H., \textit{Mathematical Theory of Transport Processes in Gases} (North--Holland, Amsterdam, 1972).

\bibitem{G03}
  Goldhirsch, I., \textit{Annu. Rev. Fluid Mech.} \textbf{35}, 267--93 (2003).

\bibitem{vNE98}
  van Noije, T. P. C., and  Ernst, M. H., \textit{Gran. Matt.} \textbf{1}, 57--64 (1998).

\bibitem{BRC96}
 Brey, J. J.,  Ruiz--Montero, M. J., and  Cubero, D., \textit{Europhys. Lett.} \textbf{48}, 359--364 (1999).

\bibitem{DB02}
 Dufty, J. W., and  Brey, J. J., \textit{J. Stat. Phys.} \textbf{109}, 433--448 (2002);  Brey, J. J., private communication.

\bibitem{GM02}
 Garz\'o, V., and  Montanero, J. M., \textit{Physica A} \textbf{313}, 336--356 (2002).

\bibitem{MG02}
 Montanero, J. M., and  Garz\'o, V., \textit{Phys. Rev. E} \textbf{67}, 021308-1--12 (2003).

\bibitem{SGD04}
 Santos, A.,  Garz\'o, V., and  Dufty, J. W., \textit{Phys. Rev. E} \textbf{69}, 061303-1--10 (2004).

\bibitem{MS96}
 Montanero, J. M.,  and  Santos, A., \textit{Phys. Rev. E} \textbf{54}, 438--444 (1996); Phys. Fluids \textbf{9}, 2057--2060 (1997).

\bibitem{GS03}
 Garz\'o, V., and  Santos, A., \textit{Kinetic Theory of Gases in Shear Flows. Nonlinear Transport} (Kluwer, Dordrecht, 2003).



\end{thebibliography}

\end{document}